\documentclass[aps,prb,showpacs,amsmath,twocolumn,amssymb,superscriptaddress,letterpaper]{revtex4}
\usepackage{mathrsfs}
\usepackage{graphicx}
\usepackage{graphics}
\usepackage{subfigure}
\usepackage{bm}
\usepackage{dcolumn}
\usepackage{amsmath,bm}
\newcommand{\nd}{{\vphantom{\dagger}}}

\usepackage{color}
\begin{document}
\title{ Tunable Anderson metal-insulator transition in quantum spin-Hall insulators }

\author{Chui-Zhen Chen}
\affiliation{ Institute of Physics, Chinese Academy of Sciences,
Beijing 100190, China}
\author{Haiwen Liu}
\affiliation{International Center for Quantum Materials, School of Physics,
Peking University, Beijing 100871, China}
\affiliation{Collaborative Innovation Center of Quantum Matter, Beijing, 100871, China}
\author{Hua Jiang}
\affiliation{College of Physics, Optoelectronics and Energy,
Soochow University, Suzhou 215006, China}
\author{Qing-feng Sun}
\affiliation{International Center for Quantum Materials, School of Physics,
Peking University, Beijing 100871, China}
\affiliation{Collaborative Innovation Center of Quantum Matter, Beijing, 100871, China}
\author{Ziqiang Wang}
\affiliation{Department of Physics, Boston College, Chestnut Hill, Massachusetts 02167, USA}
\author{X. C. Xie}
\affiliation{International Center for Quantum Materials, School of Physics,
Peking University, Beijing 100871, China}
\affiliation{Collaborative Innovation Center of Quantum Matter, Beijing, 100871, China}
\date{\today}

\begin{abstract}
We numerically study disorder effects in Bernevig-Hughes-Zhang (BHZ) model,
and find that Anderson transition  of  quantum spin-Hall insulator (QSHI)
is determined by model parameters.
The BHZ Hamiltonian is equivalent to two decoupled spin blocks that belong to the unitary class.
In contrast to the common belief that a two-dimensional unitary system scales to an insulator except at certain critical points,
we find, through calculations scaling properties of the localization length, level statistics, and
participation ratio, that  a possible exotic metallic phase emerges  between a QSHI  and a normal insulator phases in InAs/GaSb-type BHZ model.
On the other hand, direct transition from a QSHI to a
normal insulator is found in  HgTe/CdTe-type BHZ model.
Furthermore, we show that the metallic phase originates from the Berry phase and can survive both inside and outside the gap.
\end{abstract}


\maketitle

\section{ Introduction}
 Topological insulators (TI), which are identified as a class of quantum state of matter,  have generated intensive interests recently. \cite{Hasan2010,Qi2011}
The two dimensional (2D) TI, i.e., quantum spin-Hall insulator (QSHI), is characterized by odd pairs of counter propagate gapless edge states. \cite{Kane-Mele2005, BHZ2006}
The QSHI was firstly realized in HgTe/CdTe quantum well (QW)  \cite{BHZ2006,Konig2007}  and subsequently
in InAs/GaSb QW.\cite{Liu2008,Du2013} These two experimentally
realized QSHI systems can both be represented by the inverted bands
Bernevig-Hughes-Zhang (BHZ) model but with different parameters.
Specifically, the coupling strength between two inverted bands in InAs/GaSb QW is an order of magnitude smaller than that of HgTe/CdTe QW.
Considering the remarkable parameter difference, one open question remains, namely whether this difference will have some physical consequences in these two QSHI systems.

The Anderson metal-insulator transition in 2D disordered
system manifests as a lasting research issue in condensed-matter physics.\cite{Huckestein1995,Kramer1993,Belitz1994,Kramer1981,Kramer1983,Anderson1979,Mirlin2008}
Generally, the disordered electron systems can be classified into three universality ensembles
according to the random matrix theory.\cite{Beenakker1997,Mirlin2008}
In the presence of time-reversal symmetry (TRS), the system is classified as an orthogonal ensemble
if spin rotation symmetry is preserved; otherwise, it belongs to a symplectic ensemble.
In contrast, when the TRS is broken, the system turns into a unitary ensemble.\cite{Beenakker1997,Mirlin2008}
For a TRS QSHI with Rashba spin orbital coupling (SOC), the system falls into the symplectic ensemble because
spin rotation symmetry is broken. In contrast, without Rashba SOC, the QSHI is divided into two spin species of quantum anomalous
Hall (QAH) systems, which belong to the unitary ensemble. Previous studies on metal-insulator
transition in a QSHI can be summerized into two paradigms: (i) for (symplectic) QSHI with Rashba
SOC, it was found that TI and normal insulator (NI) phases are separated by a metallic
phase;\cite{Yamakage2011,Yamakage2013,Onoda2007} (ii) for a (unitary) QSHI
without Rashba SOC, a direct transition from a TI to a NI was discovered.\cite{Yamakage2011,Yamakage2013,Onoda2007,Onoda2003}
Surprisingly, recently a crossover from weak localization to weak anti-localization (WAL) is
suggested in the BHZ model without Rashba SOC.\cite{Tkachov2011,Ostrovsky2012,Krueckl2012}
Since WAL can add a positive correction to the $\beta$ function,\cite{Anderson1979,Hikami1980,Lee1985}
there is the possibility that a metallic phase might exist in the 2D unitary system (QSHI) with weak disorder, although that is against the traditional view.

\begin{figure*}[htb]
\centering
\includegraphics[scale=0.65, bb = 0 10  730 240, clip=true]{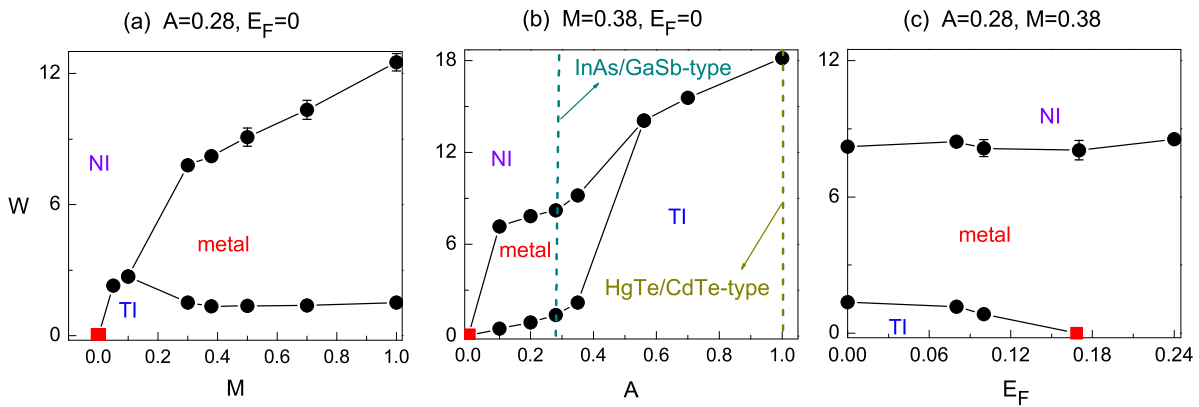}
\caption{(Color online). The phase diagrams  for disorder strength $W$ and
(a) topological mass $M$, (b) electron-hole hybrid strength $A$, and (c) Fermi energy $E_{F}$.
 The black filled circles are critical points determined by finite size scaling,
while the red filled squares for $W=0$ are analytic critical points.
%The filled circles may shift due to finite size effect.
In (b), the dark cyan dash lines (InAs/GaSb-type BHZ parameter)
indicates the  TI-metal-NI  transition.
The dark yellow dash lines (HgTe/CdTe-type parameter) shows
direct  NI-TI transition.
\label{fig4} }
\end{figure*}

In this paper, we study whether  a metallic phase can exist between TI and NI phases in a unitary  QSHI system.
Starting from the BHZ model Hamiltonian,\cite{BHZ2006,Liu2008}
we calculate  the localization length and two-terminal conductance numerically,
and we analyze the scaling behavior of the system.\cite{Datta,Kramer1983}
It is worth noting that the BHZ Hamiltonian is equivalent to two decoupled spin blocks which belong to the unitary class.
The main results  are summarized in the phase diagrams in Fig.~\ref{fig4}.
In all these phase diagrams, we find a metallic phase between TI and NI phases,
in contrast to the common view of Anderson transition behavior in a 2D unitary class.
Furthermore, we find that different parameters in BHZ model (unitary system)
can lead to different Anderson transition behaviors.
The transition from a TI to a metal is likely to exist  in InAs/GaSb-type BHZ model but not in the HgTe/CdTe-type BHZ
model [see dash lines in Fig.~\ref{fig4}(b)].
By employing the Berry phase, the parameter-dependent metallic phase can be well explained.\cite{Krueckl2012,Tkachov2013}

The rest of the paper is organized as follows. In Sec. II,
we introduce the lattice model Hamiltonian and give the details of numerical simulations.
In Sec. III, we show the main results by scaling of the localization length and the conductance.
We discuss energy level statistics and  participation ratios of eigenstates in Sec. IV.
In Sec.V, we discuss the phase diagram and interpret our numerical results by the Berry phase.
In Sec.VI, a brief summary is presented.
%The metallic phase, germinated from the WAL region, can  survive both inside and outside the band gap
%\cite{Ostrovsky2012,Krueckl2012,Tkachov2011,Imura2009,Lu2011}.

\section{Model and methods}
We consider the disorder BHZ Hamiltonian on a square lattice:\cite{BHZ2006,Liu2008,Jiang2009}
\begin{eqnarray}
H &=&
\sum_{i}\varphi_{i}^{\dagger }E_{i}\varphi_{i}^{\nd}
+\sum_{i,\alpha=x,y}\varphi _{i}^{\dagger }T_{\alpha}\varphi _{i+\widehat{\alpha}}^{\nd}+H.C.,
\end{eqnarray}
with %$E_{i}$ and $ T_{\alpha}$  are:
\begin{eqnarray}
E_{i}&=& (C-\frac{4D}{a^2}+V_{i})\sigma_{0}\otimes\tau_{0} +(M-\frac{4B}{a^2})\sigma_{0}\otimes\tau_{z},\nonumber\\
T_{x}&=& \frac{D}{a^2}\sigma_{0}\otimes\tau_{0}+\frac{B}{a^2}\sigma_{0}\otimes\tau_{z}-\frac{iA}{2}\sigma_{z}\otimes\tau_{x},\nonumber\\
T_{y}&=&\frac{D}{a^2}\sigma_{0}\otimes\tau_{0} +\frac{B}{a^2}\sigma_{0}\otimes\tau_{z}+\frac{iA}{2a}\sigma_{0}\otimes\tau_{y}.
\label{equ:2}
\end{eqnarray}
Here $i=(i_{x},i_{y})$ is the site index, and $\widehat{\alpha}$ is the unit vector along $\widehat{\alpha}=(x,y)$ direction.
$\varphi_{i}$
represents the four annihilation operators of an electron on site $i$.
The model parameters $A$, $B$, $C$, and
$D$ can be experimentally controlled
and $a$ is the lattice constant.  Specifically, two important physical parameters are coupling strength between inverted bands $A$ and
the mass $M$.  $\sigma$ and $\tau$ are  Pauli matrices in spin and orbital spaces, respectively.
We consider the  long-range disorder  potential  $V_{i}$ at  $\vec{r_{i}}$ with
$V_{i}=\sum_{n=1}^{N_{I}} U_{0}\exp[-|\vec{r_{n}}-\vec{r_{i}}|^{2}/(2\xi^2)]$,
where  $U_{0}$ is uniformly distributed in $(-W/2,W/2)$ with disorder strength $W$
and $N_{I}$ impurities are randomly located among $N$ lattices at
\{$\vec{r}_{1},\vec{r}_{2},...,\vec{r}_{N_{I}}$\}.\cite{Zhang2009,Lewenkopf2008,Rycerz2007,Wakabayashi2007}
We fix the impurity density $n=N_{I}/N=5\%$ and the disorder range $\xi=2a$,
where different $n$ and $\xi$ will not significant influence our results.
Since the two spin block are decoupled, we  only consider the spin-up block in the rest of the paper.
Because one spin  species of QSHI is a QAH insulator, our results in this paper
are applicable to the QAH or Chern Insulator.

\begin{figure*}
\centering
\includegraphics[scale=0.62, bb = 0 0 1000 300, clip=true]{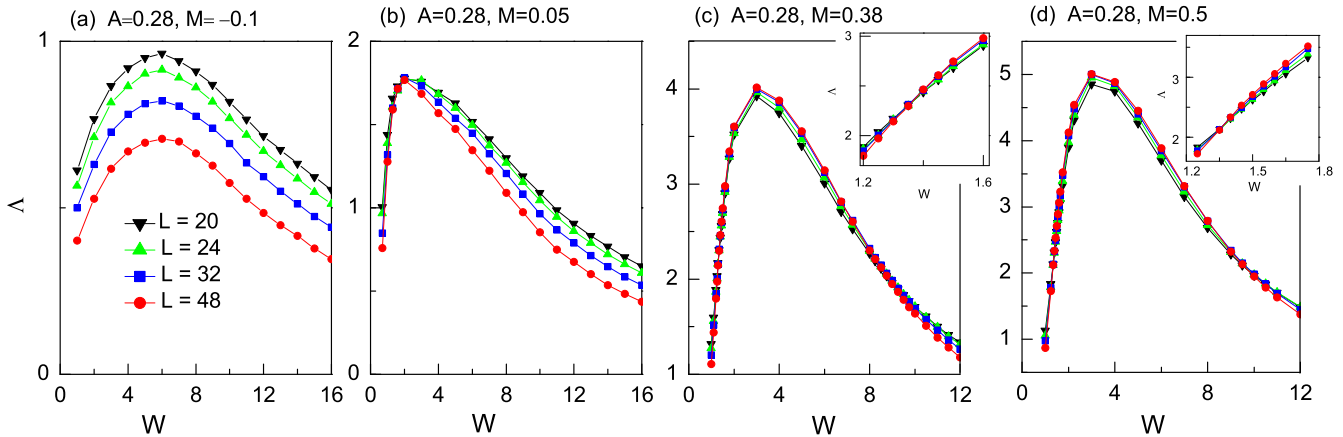}
\caption{(Color online). (a)-(d) renormalized localization length  $\Lambda=\lambda_{L}/L$ versus
disorder strength $W$ for different masses $M$ and  widths $L$ with fixed Fermi energy $E_{F}=0$.
The inset of (c) and (d) are the zoom-in of the critical point on left side of main panel of (c) and (d), respectively.}
\label{fig1}
\end{figure*}
\begin{figure}[htb]
\centering
\includegraphics[scale=0.4, bb = 0  0 650 500, clip=true]{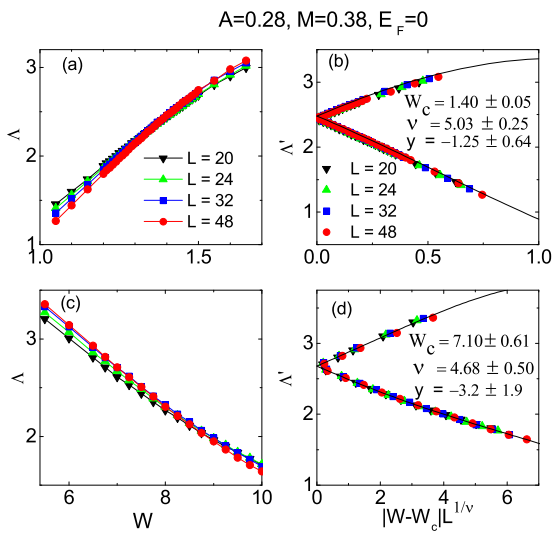}
\caption{(Color online). (a) and (c) shows
$\Lambda=\lambda_{L}/L$ vs disorder strength $W$ near two critical points.
The parameters are the same as those of Fig.~\ref{fig1}(c).
(b) and (d) are single parameter scaling of $\Lambda$ in (a) and (c), respectively.
$\nu$ is the critical exponent at critical disorder strength $W_{c}$.
\label{fig2} }
\end{figure}

In our numerical calculations, we study the localization length $\lambda_{L}$ as well as
the  dimensionless intrinsic conductance $g$ of the cylindrical sample
with width (circumference) $L$,\cite{Jiang2009} which can eliminate the effect of the helical edge states.
$g$ is defined as $1/g=1/g_{L}-1/N$, with  $g_{L}$ the two-terminal conductance
and $N$ the number of propagating channels.\cite{Braun,Zhang2009,Slevin}
The localization length $\lambda_{L}$
of the sample is calculated by transfer-matrix method \cite{Kramer1983} with the sample's length $10^{6}$ to $10^{7}$.
The two terminal conductance $g_{L}$ at Fermi energy $E_F$ is
evaluated by the Landauer-B\"{u}ttiker formula\cite{Datta,Braun,Zhang2009} with a disordered middle region with size $L \times L$ being considered as coupled to two clean semi-infinite leads.
In addition, we also investigate the energy level statistics as well as participation ratios.
For simplicity, the parameters $B = 1$, $C = 0$,
$D = 0$, and $a = 1$ are fixed in the rest of paper, where we have
assumed the particle-hole symmetry ($D = 0$).

\section{ Metal-insulator transition}
First, we study $\Lambda=\lambda_{L}/L$ versus disorder strength $W$
by increasing the mass from $M=-0.1$ to $0.5$ at a fixed  Fermi energy $E_{F}=0$, as shown in  Fig.~\ref{fig1}.
%In realistic materials, $M$ can be tuned by gate voltages.
These parameters resemble the InAs/GaSb QW $k \cdot p$ parameters.\cite{note1}
Notably, we find the TI-metal-NI transition in the InAs/GaSb-type BHZ model (unitary system)
by increasing the mass $M$.
In the NI phase with $M<0$, $\Lambda=\lambda_{L}/L$
decreases monotonously with increasing $L$ in Fig.\ref{fig1}(a)
which indicates all the states are localized.
In contrast, when $M=0.05$ in TI phase ($M>0$),
the system shows one critical (touching) point where $\Lambda$ is independent of $L$ [see Fig.\ref{fig1}(b)],
which is consistent with the previous studies of a 2D unitary system.\cite{Onoda2003,Onoda2007,Yamakage2013}
Due to the inverted gap, the TI is robust to weak disorder, and the NI phase appears
after a certain disorder strength $W_c = 2.3 \pm 0.13$.
Therefore,  this critical (touching) point indicates a direct transition from NI to TI.
Surprisingly, in Fig.~\ref{fig1}(c), when the mass $M$ is increased to $0.38$,
the metallic phase ($\Lambda$ increasing with $L$) appears between $W_{c1}\approx1.40$ % or $\ln g$
[see the inset of Fig.~\ref{fig1}(c)] and $W_{c2} \approx 7.10$.
This metallic phase is contradict to the common behavior of the 2D unitary class.
The unitary system, e.g., quantum Hall system,
is scaled to localized states except at certain critical (touching) points.
Moreover, upon further increasing $M$, the metallic phase region becomes larger, i.e., for $M=0.5$,
the metallic phase remains in Fig.~\ref{fig1}(d)
between $W_{c1}\approx1.40$ [see inset of Fig.~\ref{fig1}(d)] and $W_{c2}\approx8.2$.
This peculiar metallic phase between the TI
and NI phases in the InAs/GaSb-type BHZ model is the main finding of our paper.
%As a result, we conclude that the metallic phase
%could exist between TI and NI phases in InAs/GaSb-type
%BHZ model (unitary system). This is the main result of this paper.

Next, we analyze the one parameter scaling behavior of the renormalized localization length $\Lambda=\lambda_{L}/L$
near the critical points of Fig.~\ref{fig1}(c).
According to finite size scaling law, all  $\Lambda$ are fitted to
$\Lambda(W,L)=\Lambda_{c} + \sum^{4}_{n=1}a_{n}(W-W_{c})^{n}L^{n/\nu} + b_{0} L^{y}$ near the critical point,
\cite{Kramer1981,Kramer1983,Kramer1993,Slevin1997,Onoda2007,Obuse2007}
where $W_{c}$ is the disorder strength at a critical point, $\nu$ is  critical exponent
and $y$ is an exponent associated with the leading irrelevant operator. Here, $a_{n}$ and $b_{0}$ are the fitting parameters.
For convenience, we define
$\Lambda^{\prime}(W,L)=\Lambda(W,L)- b_{0} L^{y}$,
with the same parameters as $\Lambda$.
The best fit is given by minimizing $\chi^2$ statistic
  $\chi ^2  = \sum_{n=1}^{N} (\Lambda_{n} - \Lambda(W_{n},L_{n}))^{2}/\sigma_{n}^2$,
where $N$ is the number of the data
and $\sigma_{n}$ is the error of the $n$th data  $\Lambda_{n}$.
The finding of the metallic phase and the large critical exponent $\nu$ (almost twice as large as that obtained previously)
 in Figs.\ref{fig2}(c) and (d)
  suggests the existence of a possible new universality class.\cite{Onoda2003,Asada2002,Slevin2009,Huckestein1990}
We caution that the standard deviations  ($\chi ^2 /N $)  are $3.3$ and $3.5$ for Figs.\ref{fig2}(b) and (d), respectively,
  which are somewhat larger than the common value unity and even
larger size computations are desirable to obtain a high precision critical exponent $\nu$.

\begin{figure*}[tb]
\centering
\includegraphics[scale=0.60, bb = 0  0 1000 300, clip=true]{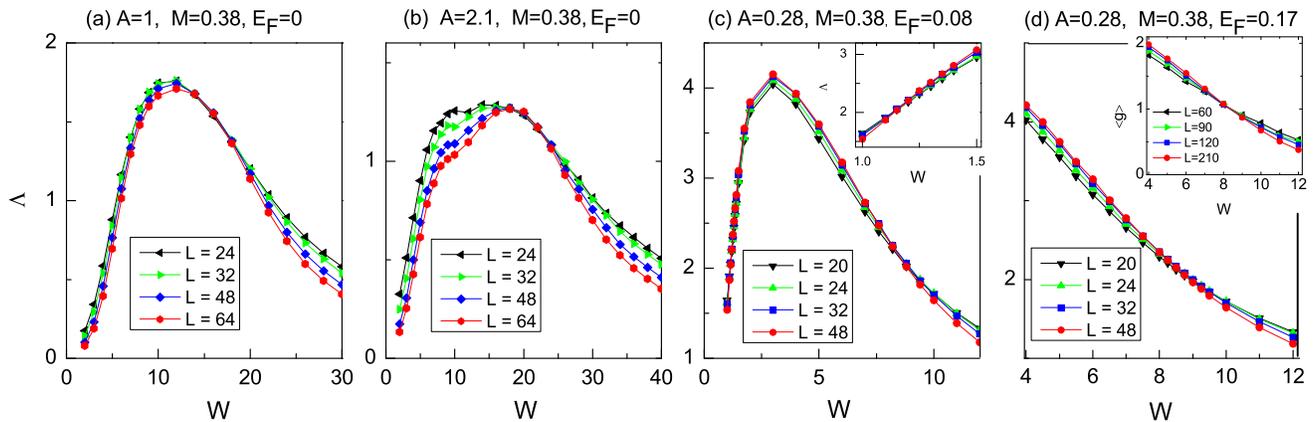}
\caption{(Color online).
(a)-(d)
$\Lambda=\lambda_{L}/L$ versus disorder strength $W$.
The parameters are shown on  top of  each panel.
The inset of (c) shows a zoom-in of the critical point on the left side in the main panel.
The inset of (d) shows average  intrinsic conductance $ \langle g \rangle$
of square sample versus  $W$ with same model parameters as  (d).}
\label{fig3}
\end{figure*}

Up to now, we have found the metallic phase for small $A=0.28$, which resembles the parameters of InAs/GaSb-type BHZ model.
To compare with previous studies,\cite{Yamakage2011,Yamakage2013} it is necessary to investigate Anderson transition
for large $A$ cases, i.e., the HgTe/CdTe-type BHZ model.
In Figs.~\ref{fig3} (a) and (b), where $M=0.38$ and $E_{F}=0$, $A$
increases from $1$ to $2.1$, in the region of the HgTe/CdTe-type model parameters.
The other parameters are the same as those of Fig.~\ref{fig1}(c).
Compared with the case $A=0.28$ [see Fig.~\ref{fig1}(c)],
the metallic phase disappears for $A=1$ and $A=2.1$
because $\Lambda$ hardly changes with the size near the touching point [see Figs.~\ref{fig3}(a) and (b)].
This direct transition from  TI to  NI (i.e. $A=~1,~2.1$) is consistent with previous study in the HgTe/CdTe-type BHZ model.\cite{Yamakage2011,Yamakage2013}
Meanwhile, the localization length decreases rapidly with increasing $A$ from $0.28$ to $2.1$.
For example, for a typical unitary case $A=1$, the localization length $\lambda_{L}$ is only $1\sim2$ times of width of the system $L$.
From the above results,  it is natural  to conclude that  the existence of the metallic phase is  highly dependent on the magnitude of $A$.
To be specific, the transition TI-metal-NI in the InAs/GaSb-type BHZ model is absent in the HgTe/CdTe-type BHZ model.

Now we consider the influence of Fermi energy $E_{F}$ on the metallic phase.
In Fig.~\ref{fig3}(c), when $A=0.28$, $M=0.38$, and  $E_{F}=0.08$ in the neighborhood of the gap center,
the transition TI-metal-NI remains almost the same as that of Fig.~\ref{fig1}(c) with $E_{F}=0$.
When the Fermi energy  is moved outside the band gap $E_{g}\simeq0.16$,  i.e.,  $E_{F}=0.17$,
the transition from metal to NI is still observed
by scaling $\Lambda$ and  $\langle g \rangle$  in Fig.~\ref{fig3}(d).
It is clear that $\langle g \rangle$ and $\Lambda$
increase with the size of the system in the metallic phase, they decrease in insulator phase,
and there is a crossing point ($W\approx8$)  at the phase transition.
To conclude, the localization length scaling
and the conductance scaling suggest that the metallic phase could survive both inside and outside the band gap.

\section{energy level statistics and participation ratio}

\begin{figure}
  \centering
  \includegraphics[scale=0.4, bb = 0 10 750 430, clip=true]{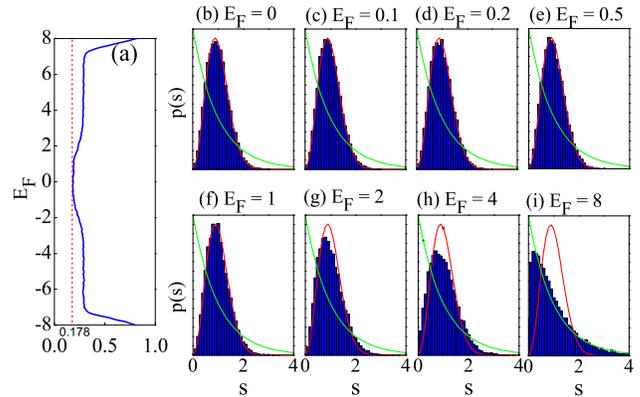}\\
  \caption{(Color online). (a) solid blue line is  the variances of the energy spacing distributions at different
   energies $E_{F}$ with the disorder strength $W=4$. The red dash line indicates the variances
   of Wigner surmise of unitary ensemble.
  (b)--(i) The histograms of nearest energy level spacings centered at different energies
  $E_{F}=0$--$8$ with $W=4$. The solid red lines and green lines represent Wigner surmise of unitary ensemble
  and Poisson distribution, respectively.  The model parameters are $A=0.28$ and $M=0.38$ with the sizes $48\times 48$.}
  \label{fig5}
\end{figure}

Furthermore, we have verified the
existence of the metallic phase by studying the energy
level statistics and evaluating the participation ratio of
eigenstates. According to the random matrix theory, the delocalized and localized states can be characterized
by the energy level statistics.\cite{Mehta,Mirlin2008}
In Fig.\ref{fig5}, the histograms of the level spacings
are drawn at different energies $E_{F}$. The sizes is $48\times48 $ with a periodical
boundary condition in two directions, i.e., a torus geometry.\cite{Prodan2010,Song2014}
In the energy region  $E_{F}=0$--$0.2$, the histograms [see Fig.\ref{fig5}(b)-(d)]
and corresponding variances [see Fig.\ref{fig5}(a)] are very close to those of the
Wigner surmise of a unitary ensemble $p(s)=(32 /\pi^{2})s^{2}\exp[-(4 /\pi) s^{2}]$.\cite{Mehta}
Therefore, the level correlation is long-ranged which indicates that extend states may exist.
However, near the top of the band ($E=8$) [see Fig.\ref{fig5}(i)], the histogram is close to the Poisson distribution
$p(s)=\exp(-s)$, which implies that the states are uncorrelated and localized.
\begin{figure*}
  \centering
  \includegraphics[scale=0.7, bb = 0 0  900 350, clip=true]{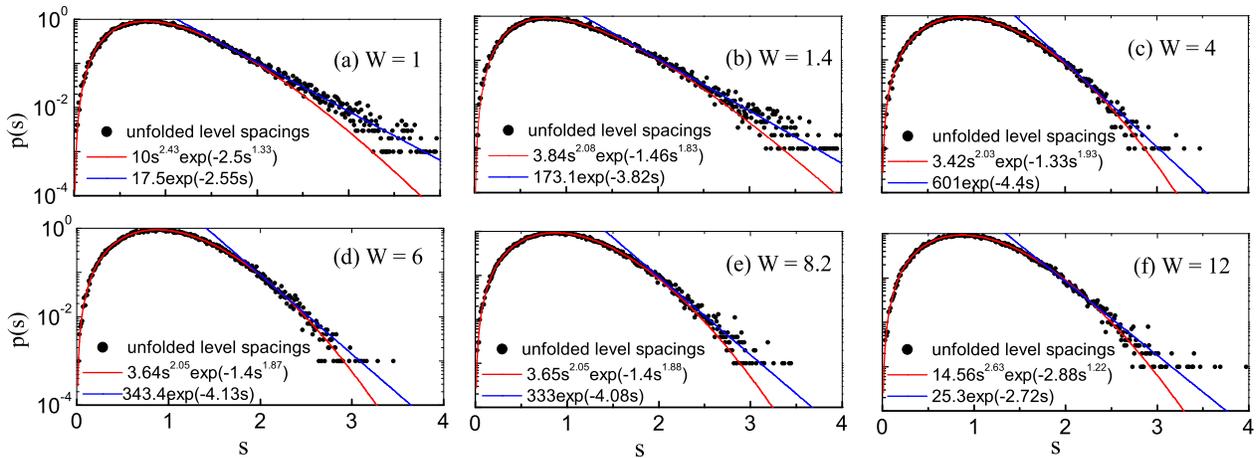}\\
  \caption{(Color online).  Distributions of  the unfolded level spacings $s$ under different disorder strength
  (a) $W=1$ to (f) $W=12$ are fitting by Wigner surmise $p(s)=As^{\alpha}\exp(Bs^{\beta})$ for whole $s$ region and Poisson distribution
  $p(s)=A \exp(Bs)$ for $s>2$.  The model parameters are $A=0.28$ and $M=0.38$ at $E_{F}=0$.}
  \label{fig6}
\end{figure*}

However, in the histograms of the level spacings, it is difficult to distinguish the critical point ( or region)
from the true metallic phase, because they are both close to Wigner surmise of the unitary class.
To identify the true metallic phase,
 we test the larger sized ($L\times L=128\times128$) results at fixed energy $E_{F}=0$
with different disorder strength $W=1$--$12$. Then we fit the distributions to
Wigner surmise $p(s)=a s^{\alpha}\exp(b s^{\beta})$ for the whole $s$ region and Poisson distribution
$p(s)=k \exp(\kappa s)$ for $s>2$, with the fitting parameter $\alpha$ ($\beta$, $\kappa$)
 and the renormalization parameter $a$ ($b$, $k$).\cite{Wang1998}
The fitting error of a set of data $\{\bar{p}(s_1),\bar{p}(s_2),...,\bar{p}(s_N)\}$ is defined as
$\sigma^2=\sum_{i=1}^{N}[p(s_n)/\bar{p}(s_n)-1]^2/N$,
where $p(s_n)$ is the fitting value at $s_n$ and $N$ is the number of data.
It is noted that if energy level spacings data follows the Wigner surmise with $\alpha=2$ and $\beta=2$, a metallic phase exits;
however, if the data only follows this distribution in the small $s$ region but violates in the large-$s$ region, the critical behavior
emerges. \cite{Shklovskii1993,Wang1998}

Figure\ref{fig6} shows a logarithmic plot of $p(s)$ for the distributions of  the unfolded level spacings $s$.\cite{footnote2}
In Figs.\ref{fig6}(c) and (d), for $W=4$ and $6$
the fitting parameters $\alpha$ and $\beta$ are very close to the Wigner surmise of the unitary ensemble,
$\alpha=\beta=2$. Therefore, the level correlation is long ranged.
On the other hand, for $W=1.4$ and $8.2$ [see Fig.\ref{fig6}(b) and (e)],
although the fitting parameters $\alpha$ and $\beta$ are very close to the Wigner surmise of the unitary ensemble,
the large $s$ region is not well described by  the Wigner surmise.
Instead, the tail of the large $s$ region is clearly fitted to Poisson distribution.
The hybrid of the Wigner surmise statistics and Poisson statistics behavior
at a critical point is coincident with the level statistics results at the 3D Anderson metal-insulator transition.\cite{Shklovskii1993}
In this case, the level correlation is finite-ranged,
which manifests as a crossing over from (long-ranged correlated) Wigner surmise statistics
to (uncorrelated) Poisson statistics. For $W=1$ and $W=12$ [see Fig.\ref{fig6}(a) and (f)],
the level spacings clearly deviated from the Wigner surmise of the unitary ensemble for the whole $s$ region
and the energy levels will become totally uncorrelated in thermodynamic limit, identifying an insulating phase.
In conclusion, the level statistics results strongly support the existence of
 a metallic phase between $W=4$ and $W=6$ at $E_{F}=0$.
This is coincident with the region indicated by finite size scaling results of localization length
(see Fig.1).

\begin{figure}
  \centering
  \includegraphics[scale=0.38, bb = 50 0 950 600, clip=true]{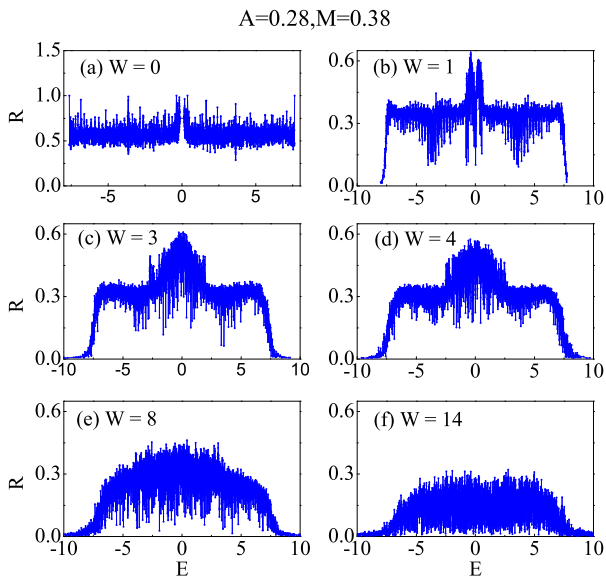}\\
  \caption{(Color online). The participation ratio R as functions of energy E
  at the disorder strength (a)$W=0$ to (f) $W=14$ at the size $48\times48$.}
  \label{fig7}
\end{figure}

The participation ratio characterized the spatial extension of the eigenstates.
Next, we will investigate the participation ratio  under different disorder strength.
The participation ratio is defined as $R= \sum_{i=1}^{N} |a_{i}|^4/(N\sum_{i=1}^{N} |a_{i}|^2)^2$
where $a_{i}$ is the wave function at lattices $i$ and $N$  is the number of the lattice. \cite{Edwards1972,Bauer1990,Zhang2009,Murphy2011}
$R$ has its maximum $1$ for one single Bloch wave, reaches a  finite value typical around $1/3$ for a disordered extended state,
and  approaches $\sim1/N$ for a localized state.
In Fig.\ref{fig7}(b) and (c), the (highly possible) extended states
characterized by the peaks $R\approx0.6$
are found near the bulk gap at $W=1$, and then come into the gap upon increasing the disorder strength $W=3$.
As a result, the metallic phase in the gap indeed comes from the extended states near the gap, which originates from
the nearly $\pi$ Berry phase.
At last, the peak of $R$  diminishes upon increasing the disorder strength [see Fig.\ref{fig7}(d)-(f)] which indicates
that the states are localized at $W=14$ ($R<1/3$).

\section{ Phase diagram and discussion}
We summarize our results  into
three phase diagrams:  $W$-$M$, $W$-$A$, and $W$-$E_{F}$, as shown in Fig.~\ref{fig4}.
In Fig.~\ref{fig4}(a),  the metallic phase emerges from TI phase
at $M \simeq 0.1$ and expands with  increasing $M$ from $0.1$ to $1$.
On the other hand, in Fig.~\ref{fig4}(b), the metallic phase exists for small A cases
and disappears for $A > 0.56$. In other words,  the transition TI-metal-NI can  exist in the InAs/GaSb-type BHZ model
but not in the HgTe/CdTe-type BHZ model [see the dash lines in Fig.\ref{fig4}(b)].
Furthermore, in Fig.~\ref{fig4}(c), the metallic phase spreads over both inside and outside the gap ($E_{g}=0.16$).
The phase diagram Fig.~\ref{fig4}(c) indicates that the  metallic phase inside and outside the gap
have the same physical origin.

In a 2D system, the metal phase is related to WAL, which adds the positive correction to
the $\beta$ function and leads to  such phase in the thermodynamic limit. \cite{Anderson1979,Lee1985,Imura2009}
In general, the WAL exists in SOC systems of a symplectic ensemble,\cite{Hikami1980}
and Dirac systems with a $\pi$ Berry phase, e.g., graphene and helical surface states of 3D TI.\cite{McCann2006,Lu2011}
In the present case, the detailed analysis of symmetry classes and Berry phases can help
us to understand the peculiar metallic phase in the inverted band BHZ model.
We first analyze the symmetry classes of the BHZ model.
The spin-$\uparrow$ part of the BHZ Hamiltonian in momentum space is
$H_{0}(k)=A (k_{x}\tau_{x}+k_{y}\tau_{y})+(M-Bk^2)\tau_{z}$,
where $A,B,M$ and $\tau$ have the same meaning as in Eq.~(\ref{equ:2}).
When the mass term $M-Bk_{F}^2=0$, the system satisfies  pseudo-time reversal symmetry,
i.e., $H_{0}(k_{F})=\tau_{y} H_{0}^{*}(-k_{F}) \tau_{y}$, which is similar to the  massless Dirac particles in graphene.
Therefore, the system is approximate symplectic and shows WAL in the region near  $M-Bk_{F}^2=0$.\cite{Tkachov2011,Ostrovsky2012}
On the contrary, when $|M-Bk_{F}^2|>> A|k_{x}+ik_{y}|$,
the two orbital bands are nearly decoupled, thus the system belongs to  orthogonal ensemble approximately
and shows weak localization.\cite{Tkachov2011,Ostrovsky2012}
Therefore, the existence of the metallic phase is dependent on the system parameters.
Moreover, the Berry phase of the BHZ model at Fermi energy $E_F$ reads: \cite{Krueckl2012,Tkachov2013}
\begin{eqnarray}
\gamma(E_{F}) = (1-\frac{(M-Bk_{F}^{2})}{\sqrt{A^{2}k_{F}^{2}+(M-Bk_{F}^{2})^{2}}})\pi,
\end{eqnarray}
where $k_F$ is the momentum at $E_F$.
The Berry phase $\gamma$ monotonously increases with $k_F$ rising from zero.
$\gamma=0$ at $k_F=0$, $\gamma=\pi$ at $k_F=\sqrt{M/B}$, and $\gamma \rightarrow 2\pi$ while $k_F$ tends to $\infty$.
Since $\gamma$ can vary from 0 to $2\pi$, the system can show both WAL and weak localization which depends on the parameters.
%
%While $\gamma$ is equal or near $\pi$, the system shows WAL and the metallic phase in the weak disorder case. Oppositely, it shows weak localization and the metallic phase disappears \cite{Krueckl2012}, while $\gamma$ near $0$ or $2\pi$.
While $M/B>>A^{2}/B^2$, $k_g=\sqrt{\frac{M}{B}-\frac{A^2}{2B^2}}\simeq \sqrt{M/B}$,
 where $k_g$ is the momentum at the conduction-band bottom and the valence-band top, and then the Berry phase $\gamma \simeq \pi$. In this case, the system exhibits WAL and the metallic phase exists between TI and NI phases (see Fig.~\ref{fig4}).
For the case of the InAs/GaSb-type BHZ model, i.e., $A=0.28$, $B=1$, and $M=0.38$, $M/B>>A^{2}/B^2$ is well satisfied, the metallic phase consequently appears.
In contrast,
for the HgTe/CdTe type model, i.e., $A=1$, $B=1$, and $M=0.38$ with $M/B<A^{2}/B^2$, $k_g=0$ which is far away from $\sqrt{M/B}$. At $k_g=0$, the Berry phase $\gamma =0$, which leads the weak localization behavior and the disappearance of the metallic phase.

%\section{Summary}
\section{Discussion and Conclusion}
The Berry phase argument is well applicable to the present numerical simulations
and it is also in accordance with previous investigations. \cite{Tkachov2011,Ostrovsky2012}
When the system is ``approximate symplectic", electrons interfere as a symplectic system on the
length scalings smaller than $l_{M}=v_{F}\tau_{M}$. Here $v_{F}$ is the Fermi velocity,
$\tau_{M}= \frac{\tau}{2}(A^{2}k_{F}^{2}+(M-Bk_{F}^{2})^{2})/(M-Bk_{F}^{2})^2$ is the TRS-breaking scattering time
and  $\tau$ is the elastic scattering time.\cite{Tkachov2011,Ostrovsky2012}
In this case, $l_M$ acts as large-size cufoff for WAL correction resembling the dephasing length,
and the WAL correction dominates if system width $L<<l_M$.\cite{Tkachov2011,Ostrovsky2012}
In our numerical simulation, for example, when $E_{F}=0.17$ in Fig.4(d), $l_M=v_{F}\tau_{M}=l_s\tau_{M}/\tau\sim300$,
where the mean free path $l_s\sim \sqrt{1/n}\sim 4$ and $\tau_{M}/\tau \sim 75$.
Here we have used the impurity density $n=5\%$ and the Fermi vector $k_{F}\sim0.605$. Since $l_M >> L$ the system width in the numerical simulations, the TRS-breaking scattering time
$\tau_{M}$ is not important and the system resembles symplectic with metallic behaviors.
This ¡¤¡¤metallic¡° behavior can show up in mesoscopic systems\cite{Pikulin2014} and may apply to recent transport experiments
\cite{Du2013}.
For larger systems with $L>L_{M}$, the scaling behavior should be interesting, however this is beyond our present numerical capability, and thus is left for further study.

In summary, we investigated the Anderson metal-insulator transition in QSHI
and found different localization behaviors depending on model parameters.
Notably, the transition TI-metal-NI likely exists
in InAs/GaSb-type systems but not in HgTe/CdTe-type systems.
The peculiar metallic phase, which originates from the Berry phase $\pi$ near the band gap,
contradicts the common view of the Anderson transition behavior of the 2D unitary class.

\section{Acknowledgements}
 This work was financially supported by
NBRP of China (2015CB921102,2012CB921303, 2012CB821402, and 2014CB921901) and NSF-China under
Grants Nos. 11274364, NO. 91221302 and No. 11374219. ZW is supported by
DOE Basic Energy Sciences grant DE-FG02-99ER45747. H.J. is supported by the NSF of Jiangsu province BK20130283.


\begin{thebibliography}{localization in InAs/GaSb}

\bibitem{Hasan2010} M. Z. Hasan and C. L. Kane, Rev. Mod. Phys. {\bf82}, 3045 (2010).
\bibitem{Qi2011} X.-L. Qi and S.-C. Zhang, Rev. Mod. Phys. {\bf83}, 1057 (2011).
\bibitem{Kane-Mele2005}  C. L. Kane and E. J. Mele, Phys. Rev. Lett. {\bf95}, 226801 (2005).
\bibitem{BHZ2006} A. Bernevig, T. Hughes, and S. C. Zhang, Science {\bf 314}, 1757 (2006).
\bibitem{Konig2007}  M. K\"{o}nig, S. Wiedmann, C. Br\"{u}ne, A. Roth, H. Buhmann,
  L. Molenkamp, X.-L. Qi, and S.-C. Zhang, Science {\bf 318}, 766 (2007).
\bibitem{Liu2008} C. Liu, T. L. Hughes, X.-L. Qi, K. Wang, and S.-C. Zhang, Phys. Rev. Lett. {\bf100}, 236601 (2008).

\bibitem{Du2013} L. Du, I. Knez, G. Sullivan, and R.-R. Du, Phys. Rev. Lett. {\bf114}, 096802 (2015).




\bibitem{Anderson1979}E. Abrahams, P. W. Anderson, D. C. Licciardello,
and T. V. Ramakrishnan, Phys. Rev. Lett. {\bf42}, 673 (1979).
\bibitem{Kramer1981}A. MacKinnon and B. Kramer, Phys. Rev. Lett. {\bf47}, 1546 (1981).
\bibitem{Kramer1983}A. MacKinnon and B. Kramer, Z. Phys. B {\bf53}, 1 (1983).
\bibitem{Kramer1993} B. Kramer and A. MacKinnon, Rep. Prog. Phys. {\bf56}, 1469 (1993)
\bibitem{Belitz1994}D. Belitz and T. R. Kirkpatrick, Rev. Mod. Phys. {\bf66}, 261 (1994).
\bibitem{Huckestein1995} B. Huckestein, Rev. Mod. Phys. {\bf67}, 357 (1995).
\bibitem{Mirlin2008} F. Evers and A. D. Mirlin, Rev. Mod. Phys. {\bf80}, 1355 (2008).
\bibitem{Beenakker1997} C. W. J. Beenakker, Rev. Mod. Phys. {\bf69}, 731 (1997).


\bibitem{Onoda2007} M. Onoda, Y. Avishai, and N. Nagaosa, Phys. Rev. Lett. {\bf98}, 076802 (2007).
\bibitem{Yamakage2011}A. Yamakage, K. Nomura, K. I. Imura, and Y. Kuramoto, J. Phys. Soc. Jpn. {\bf80}, 053703 (2011).
\bibitem{Yamakage2013}  A. Yamakage, K. Nomura, K.-I. Imura, and Y. Kuramoto, Phys. Rev. B {\bf87}, 205141 (2013).
\bibitem{Onoda2003}M. Onoda and N. Nagaosa, Phys. Rev. Lett. {\bf90}, 206601 (2003).
\bibitem{Tkachov2011}G. Tkachov and E. M. Hankiewicz, Phys. Rev. B {\bf84}, 035444 (2011).
\bibitem{Ostrovsky2012}P. M. Ostrovsky, I. V. Gornyi, and A. D. Mirlin, Phys. Rev. B {\bf86}, 125323 (2012).
\bibitem{Krueckl2012}V. Krueckl and K. Richter, Semicond. Sci. Technol. {\bf27}, 124006 (2012).
\bibitem{Hikami1980}S. Hikami, A. I. Larkin, and Y. Nagaoka, Prog. Theor. Phys. {\bf63}, 707 (1980).
\bibitem{Lee1985} P. A. Lee and T. V. Ramakrishnan, Rev. Mod. Phys. {\bf57}, 287 (1985).


\bibitem{Datta} S. Datta, Electronic Transport in Mesoscopic Systems
(Canmbridge University Press, Cambridge, England, 1995).



\bibitem{Tkachov2013} G. Tkachov, Phys. Rev. B {\bf88}, 205404 (2013).


\bibitem{Jiang2009} H. Jiang, L. Wang, Q.-F. Sun, and X. C. Xie, Phys. Rev. B {\bf80}, 165316 (2009).
                    J.-C. Chen, J. Wang, and Q.-F. Sun, ibid. {\bf85}, 125401 (2012).


\bibitem{Rycerz2007} A. Rycerz, J. Tworzyd{\l}o, and C. W. J. Beenakker, Europhys. Lett. {\bf79}, 57003 (2007).
\bibitem{Wakabayashi2007}K. Wakabayashi, Y. Takane, and M. Sigrist, Phys. Rev. Lett. {\bf99}, 036601 (2007).
\bibitem{Lewenkopf2008}C. H. Lewenkopf, E. R. Mucciolo, and A. H. Castro Neto, Phys. Rev. B {\bf77}, 081410(R) (2008).
\bibitem{Zhang2009} Y.-Y. Zhang, J.-P. Hu, B. A. Bernevig, X. R. Wang, X. C. Xie, and W. M. Liu, Phys. Rev. Lett. {\bf102}, 106401 (2009).

\bibitem{Braun} D. Braun, E. Hofstetter, G. Montambaux, and A. MacKinnon, Phys. Rev. B {\bf55}, 7557 (1997).
\bibitem{Slevin} K. Slevin, P. Marko\v{s}, and T. Ohtsuki, Phys. Rev. Lett. {\bf86}, 3594 (2001).

\bibitem{note1} The dimensionless parameters $A=0.28$, $B=1$, $C=0, D=0, M=-0.1$ to $0.5$
correspond to  $A=37$meV$\cdot$nm, $B=-660.8$meV$\cdot$nm, C=0, $D=0$ and $M=2.6$meV to $-13.2$meV,
with the lattice constant $a=5$nm. They are InAs/GaSb-type $k \cdot p$ parameters because
$A$ is about an order of magnitude smaller than that of HgTe/CdTe-type.





\bibitem{Slevin1997}K. Slevin and T. Ohtsuki, Phys. Rev. Lett. {\bf78}, 4083 (1997).
\bibitem{Obuse2007} H. Obuse, A. Furusaki, S. Ryu, and C. Mudry, Phys. Rev. B {\bf76}, 075301 (2007).




\bibitem{Huckestein1990} B. Huckestein and B. Kramer, Phys. Rev. Lett. {\bf64}, 1437 (1990).
\bibitem{Asada2002} Y. Asada, K. Slevin, and T. Ohtsuki, Phys. Rev. Lett. {\bf89}, 256601 (2002).
\bibitem{Slevin2009} K. Slevin and T. Ohtsuki, Phys. Rev. B {\bf80}, 041304(R) (2009).

\bibitem{Mehta} M. L. Mehta, Random Matrices, 2nd ed. (Academic, Boston, 1991).
\bibitem{Prodan2010} E. Prodan, T. L. Hughes, and B. A. Bernevig, Phys. Rev. Lett. {\bf105}, 115501 (2010).
\bibitem{Song2014} J. Song, C. Fine, and E. Prodan, Phys. Rev. B {\bf90}, 184201 (2014).
\bibitem{Wang1998} V. Plerou and Z. Wang, Phys. Rev. B {\bf58}, 1967  (1998).
\bibitem{Shklovskii1993} B. I. Shklovskii, B. Shapiro, B. R. Sears, P. Lambrianides, and H.
B. Shore, Phys. Rev. B {\bf47}, 11487 (1993).
\bibitem{footnote2} The logarithmic plots (see Fig.~\ref{fig6}) show strongly scattered data in large $s$ may be misleading.
In fact,Y-axis are logarithmic, and thus the deviation in large $s$ data are small. For example, in Fig.\ref{fig6},
the error $\sigma$ of the Wigner surmise (red line) in large $s$ region are $3\sim 8\times10^{-3}$.
The fitting parameter are mainly determinted by the large $p(s)$ data by minimize the error $\sigma^2$.




\bibitem {Edwards1972} J. T. Edwards and D. J. Thouless, J. Phys. C {\bf5}, 807 (1972).
\bibitem {Bauer1990} J. Bauer, T. M. Chang and J. L. Skinner, Phys. Rev. B {\bf42}, 8121 (1990).
\bibitem {Murphy2011} N. C. Murphy, R. Wortis, and W. A. Atkinson, Phys. Rev. B {\bf83}, 184206 (2011).


\bibitem{Imura2009} K.-I. Imura, Y. Kuramoto, and K. Nomura, Phys. Rev. B {\bf80}, 085119 (2009).
\bibitem{McCann2006}  H. Suzuura and T. Ando, Phys. Rev. Lett. {\bf89}, 266603 (2002).
E. McCann, K. Kechedzhi, V. I. Fal'ko, H. Suzuura, T. Ando, and B. L. Altshuler, ibid. {\bf97}, 146805 (2006).
\bibitem {Lu2011} H.-Z. Lu, J. Shi, and S.-Q. Shen, Phys. Rev. Lett. {\bf107}, 076801 (2011).
\bibitem{Pikulin2014} D. I. Pikulin, T. Hyart, Shuo Mi, J. Tworzyd{\l}o, M. Wimmer, and C. W. J. Beenakker, Phys. Rev. B {\bf89}, 161403(R) (2014).


\end{thebibliography}
\end{document}